\documentstyle[12pt,epsf]{article}
\catcode`\@=11

\begin{document}
\begin{titlepage}
\samepage{
\setcounter{page}{1}
\rightline{JHU-TIPAC-97004}
\rightline{February, 1997}
\vfill
\vfill
\begin{center}
 {\Large \bf A Model with Anomalous U(1) Induced Supersymmetry Breaking\\}
\vfill
\vfill
 {\large Ren-Jie Zhang \footnote{E-mail address: zhang@dirac.pha.jhu.edu} \\ }
\vspace{.25in}
 {\it  Department of Physics and Astronomy\\
       The Johns Hopkins University\\
       Baltimore, Maryland 21218 USA\\}
\end{center}
\vfill
\vfill
\begin{abstract}
  {\rm A model based on an anomalous Abelian symmetry 
$U(1)_1\times U(1)_2$ is presented.
This symmetry is responsible for both supersymmetry breaking and
fermion mass generation. Quark and squark mass matrices 
are aligned to prevent large flavor-changing neutral current
and CP-violation.}
\end{abstract}
\vfill}
\end{titlepage}


\catcode`@=11
\long\def\@caption#1[#2]#3{\par\addcontentsline{\csname
  ext@#1\endcsname}{#1}{\protect\numberline{\csname
  the#1\endcsname}{\ignorespaces #2}}\begingroup
    \small
    \@parboxrestore
    \@makecaption{\csname fnum@#1\endcsname}{\ignorespaces #3}\par
  \endgroup}
\catcode`@=12

\newcommand{\newc}{\newcommand}
\newc{\gsim}{\lower.7ex\hbox{$\;\stackrel{\textstyle>}{\sim}\;$}}
\newc{\lsim}{\lower.7ex\hbox{$\;\stackrel{\textstyle<}{\sim}\;$}}
\def\tr{\mathop{\rm tr}}
\def\Tr{\mathop{\rm Tr}}
\def\Im{\mathop{\rm Im}}
\def\Re{\mathop{\rm Re}}
\def\bR{\mathop{\bf R}}
\def\bC{\mathop{\bf C}}
\def\lie{\mathop{\hbox{\it\$}}} 
\newc{\Qeff}{Q_{\rm eff}}
\newc{\mz}{M_Z}
\newc{\mpl}{M_{\rm pl}}
\newc{\vphi}{\varphi}
\newc{\ol}{\overline}

\newc{\npb}{{\it Nucl. Phys. B\ }}
\newc{\plb}{{\it Phys. Lett. B\ }}
\newc{\prd}{{\it Phys. Rev. D\ }}
\newc{\prl}{{\it Phys. Rev. Lett.\ }}
\newc{\pr}{{\it Phys. Rep.\ }}
%




\noindent {\bf 1}. Supersymmetry is a good candidate
for a theory beyond the standard model, and 
several supersymmetric extensions of the standard model
have been constructed \cite{rev}. 
There are usually two sectors in these models:
a hidden sector, in which supersymmetry is broken;
and an observable sector, which contains the standard model 
fields and their supersymmetric partners. 
Supersymmetry breaking is transmitted to the observable world either
by gravitational interactions 
or by standard model gauge interactions. 
The soft parameters in the effective Lagrangian for
the observable sector are subject to severe constraints 
from low energy rare processes. For example, the inter-generation
scalar-quark masses should be highly suppressed to avoid 
large flavor-changing neutral currents.

Recently, another scenario for supersymmetry breaking has been 
proposed \cite{Dva1, BiD}, in which supersymmetry breaking is 
triggered by an anomalous $U(1)$ symmetry which may
arise from string compactification \cite{Dterm}. 
In the simplest model \cite{Dva1}, 
the $D$-term contribution to the scalar-quark mass 
depends on the $U(1)$ charges, and dominates
over the supergravity-mediated $F$-term contribution. The first two
generation scalar-quarks are assigned the same charges and thus
have degenerate masses. Because of this, the contributions
to flavor-changing neutral current and CP-violating process are
suppressed.

Anomalous $U(1)$ symmetries have also been widely used as horizontal
symmetries for flavor physics \cite{U1A}. The breaking of
the horizontal symmetry 
gives a small expansion parameter which can be identified as 
the Cabibo angle. In this paper, we will present a model in which 
supersymmetry breaking is induced by an
anomalous $U(1)_1\times U(1)_2$ symmetry. This symmetry 
also serves as a horizontal flavor symmetry, 
and reproduces the right fermion masses and CKM matrix.
Flavor-changing neutral currents are suppressed because of
the quark-squark mass alignment mechanism \cite{Nir}.


\noindent {\bf 2}.
The model presented here is similar to that of 
Ref. \cite{Dva1}, which has an anomalous $U(1)$ symmetry
under which the supersymmetric standard model superfields $\Phi_i$ 
carry nonnegative charges $q_{\Phi_i}$. In addition, there are
two standard model singlet chiral superfields $\phi_+$ and $\phi_-$ with  
charges $+1$ and $-1$ respectively. The cubic anomaly and
mixed gauge and gravitational 
anomalies are cancelled by a 4-D analogue of
the Green-Schwarz mechanism \cite{GS}, which requires
\begin{equation}
C_a/k_a = {\Tr q_{\Phi_i}^3}/3 
= {\Tr q_{\Phi_i}}/12 = 16 \pi^2 \delta_{\rm GS},
\end{equation}
where the $C_a$'s are coefficients for the mixed gauge anomalies, 
$C_a = 2 \sum_{\Phi_i} T_a(R_{\Phi_i}) q_{\Phi_i}$,
and $T_a(R_{\Phi_i})$ is the
index of the representation $R_{\Phi_i}$. The $k_a$'s are  
the Kac-Moody levels for the gauge group: 
$1,1,5/3$ for $SU(3),\ SU(2),\ U_Y(1)$ respectively.
The Green-Schwarz mechanism induces a Fayet-Iliopoulos 
$D$-term $\xi D$ \cite{Dterm}, where 
\begin{equation}
\xi = g^2\delta_{\rm GS} M^2_{\rm pl} = \lambda^2 M^2_{\rm pl},
\end{equation}
$\lambda = g \delta_{\rm GS}^{1/2}$, and $g$ is the 
string coupling constant. 

In the model of Ref. \cite{Dva1}, the effective scalar potential is 
\footnote{The mass parameter $m$ is assumed 
to be generated dynamically from the hidden sector
\cite{Dva1}.}
\begin{equation}
V\ =\ 
{g^2\over 2}\biggl[\sum_i q_{\Phi_i} |\Phi_i|^2 + |\phi_+|^2 
- |\phi_-|^2 + \xi + \cdots\biggr]^2 + m^2 |\phi_+|^2 + m^2 |\phi_-|^2,
\end{equation}
where we have omitted $D$-term contributions from the
hidden sector fields that are charged under the anomalous $U(1)$.
This potential
gives rise to supersymmetry and anomalous $U(1)$ symmetry breaking.
The nonvanishing vevs are
\begin{equation}
<\phi_->\ \simeq\ \lambda M_{\rm pl}, ~~< F_{\phi_+}>\ \simeq\ \lambda m \mpl,
~~ <D>\ =\ {m^2\over g^2}\ .
\end{equation}

In this model, 
the soft scalar masses contain $D$-term contributions, which 
depend on the $U(1)$ charges. They also receive 
supergravity-mediated $F$-term contributions, as follows,
\begin{equation}
{\tilde m}^2_{\Phi_i}\ \simeq\ q_{\Phi_i} m^2 + {F^2_{\phi_+}\over \mpl^2} 
\ \simeq\ (q_{\Phi_i} + \lambda^2) m^2\ .
\label{eq:mass}
\end{equation}
Note that for a $U(1)$ charged field, the $D$-term contribution 
dominates over the $F$-term contribution.

The gaugino masses are generated by a term
$\phi_+\phi_- W_a W_a/M^2_{\rm pl}$ in the superpotential. They are 
\begin{equation}
m_\lambda\ \simeq\ {<F_{\phi_+}\phi_->\over M^2_{\rm pl}}\ \simeq\ \lambda^2 m.
\end{equation}
The experimental lower limit for gaugino masses requires 
$m$ of order a few TeV for $\lambda \simeq 0.2$.

If the anomalous $U(1)$ symmetry is used as a 
horizontal symmetry for fermion mass generation \cite{Moha},
it generically does not give
the correct CKM matrix. Although the result is general, 
we will show it for the case 
where supersymmetric Higgs fields 
$H_1$ and $H_2$ have zero $U(1)$ charges.
(Negative $U(1)$ charges  would
result in Planck scale Higgs vevs. Positive $U(1)$ charges 
can give rise to a Planck scale $\mu$-term. Even if this
$\mu$-term vanishes, the Higgs particles
receive TeV-scale soft masses from eq. (\ref{eq:mass}),  
which reintroduces the fine tuning problem.) 
 
To ensure the proper pattern of symmetry breaking, 
we must assign nonnegative $U(1)$ charges
to all quark and lepton superfields. 
Then the Yukawa couplings are 
given by the following terms in the
superpotential, 
\begin{equation} 
\biggl({\phi_-\over\mpl}\biggr)^{q_{Q_i}+q_{{\ol U}_j}} Q_i H_2 {\ol U_j}
+\biggl({\phi_-\over\mpl}\biggr)^{q_{Q_i}+q_{{\ol D}_j}} Q_i H_1 {\ol D_j}\ .
\label{eq:yukawa}
\end{equation}
If we assign the same $U(1)$
charges to the first two generation superfields to suppress FCNC,
from eq. (\ref{eq:yukawa}) we find   
the following relations among Yukawa matrix entries, 
$Y^u_{11} \simeq Y^u_{12} \simeq Y^u_{21} \simeq Y^u_{22}$ 
and similarly $Y^d_{11} \simeq Y^d_{12} \simeq Y^d_{21} \simeq Y^d_{22}$.
Such Yukawa entries do not generically give a satisfactory CKM matrix.

The other way to suppress FCNC is to align the squark and quark 
mass matrices. This permit us to assign 
different $U(1)$ charges to the first two generation 
superfields.
In our case, however, 
we find $Y^d_{11}/Y^d_{12} \simeq Y^d_{21}/Y^d_{22}$,
independent of the $U(1)$ charges assigned,
which is not compatible
with the alignment restriction that both $Y^d_{21}$ and
$Y^d_{12}$ must be sufficiently small \cite{Nir}. An extra
symmetry is required to ensure ``texture zeroes" for the appropriate
Yukawa matrix entries. 

\noindent {\bf 3}.
In the rest of this Letter we consider a model 
with two anomalous $U(1)$ symmetries. We take
the standard model superfields 
to have the following charges under this $U(1)_1\times
U(1)_2$ symmetry\footnote{The mixed gauge anomalies are cancelled
by the Green-Schwarz mechanism under the above charge assignment, assuming 
there is no extra standard model and anomalous $U(1)$ charged matter.
The hidden sector fields carry the anomalous $U(1)$ charges as well, 
to ensure the cancellation of the mixed gravitational anomaly and 
the mixed gauge anomaly for the hidden sector gauge group.} 

\begin{equation}
\begin{array}{c|ccccccccccccccccc}
 & Q_1 & Q_2 & Q_3 & {\ol U}_1 & {\ol U}_2 & {\ol U}_3
& {\ol D}_1 & {\ol D}_2 & {\ol D}_3 & L_1 & L_2 & L_3 & 
{\ol E}_1 & {\ol E}_2 & {\ol E}_3 & H_1 & H_2 \\
\hline
q_1 & 6 & 2 & 0 
    & 0 & -2 & 0
    &-6 & 2 & 2
    &-5 & -7& 0
    & 5 & 11& 2
    & 0 & 0\\
q_2 & -3& 0 & 0
    & 3 & 3 & 0 
    & 9 & 0 & 0
    & 8 & 10& 0
    &-2&-10& 0
    &0 & 0 \\
\end{array} . 
\end{equation}
The charges for standard model singlets are 
\begin{equation}
\begin{array}{c|cccc}
    & \phi_+ & \phi_- & \psi_+ & \psi_- \\
\hline
q_1 &      1 &     -1 &      0 &      0 \\
q_2 &      0 &      0 &      1 &     -1  
\end{array}\ .
\end{equation}
Note that some fields have negative charges under one $U(1)$
and positive charges under the other, but that the sums 
of these charges are always 
positive.
A slightly different model has been considered before \cite{Nir, LNS} 
in a different context.

With these assumptions, the scalar potential of this model is as follows,
\begin{eqnarray}
V & = & 
{g^2\over 2}\Biggl\{\biggl[\sum_i q_{\Phi_i1} |\Phi_i|^2 + |\phi_+|^2 
- |\phi_-|^2 + \xi_1\biggr]^2 +
\biggl[\sum_i q_{\Phi_i2} |\Phi_i|^2 + |\psi_+|^2 
- |\psi_-|^2 + \xi_2\biggr]^2\Biggr\} \nonumber\\
& & +\ m_1^2 \biggl(|\phi_+|^2 + |\phi_-|^2\biggr)
+ m_2^2 \biggl(|\psi_+|^2 + |\psi_-|^2\biggr),
\end{eqnarray}
where parameters $\xi_1, \xi_2$ are defined as before, and 
the mass parameters $m_{1,2}$ are assumed to be generated dynamically
from the hidden sector.

This potential gives rise to supersymmetry and anomalous 
$U(1)$ symmetry breaking as before, the nonvanishing vevs are
\begin{eqnarray}
&&<\phi_->\ \simeq\ \lambda_1\mpl, 
~~~<\psi_->\ \simeq\ \lambda_2\mpl,\nonumber\\
&&
< F_{\phi_+}>\ \simeq\ \lambda_1 m_1\mpl,
~~ <F_{\psi_+}>\ \simeq\ \lambda_2 m_2\mpl\ ,\label{eq:vev}\\
&&<D_1>\ =\ {m_1^2\over g^2}, ~~~
<D_2>\ =\ {m_2^2\over g^2}\ ,\nonumber 
\end{eqnarray} 
where $\lambda_i = g \delta^i_{\rm GS}$, $i=1,2$.
(For the above charge assignment, $C^1_{SU(3)} = 12, C^2_{SU(3)}= 9$, 
and $\lambda_1 \simeq \lambda_2 \simeq 0.2$.)
Note that the negative $U(1)_1$ or
$U(1)_2$ charged fields do not give rise to  
minimum which breaks the 
standard model gauge groups. 
To see this, we consider a standard model field $\Phi$
with $q_1 > 0$,
$q_2 < 0$ and $q_1+q_2 >0$.
The $D$-flat direction 
corresponds to 
\begin{eqnarray}
&&<|\phi_-|^2> = \xi_1 + q_1 <|\Phi |^2> + <|\phi_+|^2>, 
\nonumber\\
&&<|\psi_-|^2> = \xi_2 + q_2 <|\Phi |^2> + <|\psi_+|^2>.
\end{eqnarray}
It is easy to see that along this direction, the minimum of
the scalar potential is 
at $<\Phi> = 0$ and eq. (\ref{eq:vev}). 

Under the above charge assignment, the superpotential 
contains additional 
terms\footnote{For the case $q_{Q_ii}+q_{{\ol U}_ji} < 0$,  
renormalization of the K\"ahler potential induces
similar terms.}
\begin{equation}
(\phi_-/M_{\rm pl})^{q_{Q_i1}+q_{{\ol U}_j1}} 
(\psi_-/M_{\rm pl})^{q_{Q_i2}+q_{{\ol U}_j2}}
Q_i H_2 {\ol U}_j \biggl(1 + {\phi_-\phi_+\over\mpl^2} +
{\psi_-\psi_+\over\mpl^2}+\cdots\biggr)\ .
\label{pot}
\end{equation}
where $q_{Q_ii}+q_{{\ol U}_ji}\geq 0$, $i=1,2$.
The dots represent terms suppressed by more powers of $\mpl$.
The up-type quark Yukawa matrix has the following form 
(assuming $\lambda_1\simeq\lambda_2\simeq\lambda$, 
where $\lambda\simeq 0.2$
is the Cabibo angle),
\begin{equation}
Y^u\simeq 
\left(\begin{array}{ccc}
\lambda^6 & \lambda^4 & \lambda^9\\
\lambda^5 & \lambda^3 & \lambda^2\\
\lambda^3 & \lambda^5 & 1         
\end{array}\right).
\end{equation}
Similary, the down-type quark and 
lepton Yukawa matrices are given by
\begin{equation}
Y^d\simeq 
\left(\begin{array}{ccc}
\lambda^6 & \lambda^{11} & \lambda^{11}\\
\lambda^{13} & \lambda^4 & \lambda^4\\
\lambda^{15} & \lambda^2 & \lambda^2         
\end{array}\right)\ ,
~~
Y^e\simeq 
\left(\begin{array}{ccc}
\lambda^6 & \lambda^8 & \lambda^{11}\\
\lambda^{10} & \lambda^4 & \lambda^{15}\\
\lambda^7 & \lambda^{21} & \lambda^2         
\end{array}\right).
\end{equation}
It is easy to check that the fermions 
obey the following mass relations
at the $U(1)$-breaking scale, $\lambda\mpl$,
\begin{equation}
m_u : m_c : m_t \sim \lambda^6 : \lambda^3 : 1,
~~
m_d : m_s : m_b\ \ {\rm and}\  \
m_e : m_\mu : m_\tau  \sim \lambda^4 : \lambda^2 : 1.
\end{equation}
Note that $\lambda_b/\lambda_t \simeq \lambda^2$, which
implies $\tan\beta$ is small ($\sim 2$).

This fermion Yukawa texture also 
gives the correct order in $\lambda$ for 
CKM matrix elements,
\begin{equation}
|V_{ud}|, |V_{cs}|, |V_{tb}| \sim 1, ~~ |V_{us}|, |V_{cd}| \sim \lambda,
~~
|V_{cb}|, |V_{ts}| \sim \lambda^2, ~~
|V_{ub}|, |V_{td}| \sim \lambda^3\ .
\end{equation}

The soft scalar masses are determined as before,
\begin{equation}
{\tilde m_{\Phi_i}}^2 
\simeq (q_{\Phi_i1}+\lambda_1^2) m_1^2 
+  (q_{\Phi_i2} +\lambda_2^2) m_2^2.
\end{equation}
Taking $m_1\simeq m_2\simeq m$, we find the following
scalar mass matrices
\begin{eqnarray}
&&
{({\tilde m^2}_{Q})_{LL}\over m^2} \simeq
\left(\begin{array}{ccc}
3 & \lambda^7 & \lambda^9\\
\lambda^7 & 2 & \lambda^2\\
\lambda^9 & \lambda^2 & \lambda^2
\end{array}\right ) , \nonumber\\
&&
{({\tilde m^2}_{\ol U})_{RR}\over m^2} \simeq
\left(\begin{array}{ccc}
3 & \lambda^2 & \lambda^3\\
\lambda^2 & 1 & \lambda^5\\
\lambda^3 & \lambda^5 & \lambda^2
\end{array}\right ) , ~~
{({\tilde m^2}_{\ol D})_{RR}\over m^2} \simeq
\left(\begin{array}{ccc}
3 & \lambda^{17} & \lambda^{17}\\
\lambda^{17} & 2 & 0\\
\lambda^{17} & 0 & 2
\end{array}\right )\ ,\\
&&
{({\tilde m^2}_{L})_{LL}\over m^2} \simeq
\left(\begin{array}{ccc}
3 & \lambda^{4} & \lambda^{13}\\
\lambda^{4} & 3 & \lambda^{17}\\
\lambda^{13} & \lambda^{17} & \lambda^2
\end{array}\right ), ~~
{({\tilde m^2}_{\ol E})_{RR}\over m^2} \simeq
\left(\begin{array}{ccc}
3 & \lambda^{14} & \lambda^5\\
\lambda^{14} & 1 & \lambda^{19}\\
\lambda^5 & \lambda^{19} & 2 \nonumber
\end{array}\right )\ ,
\label{eq:squark}
\end{eqnarray}
where the off-diagonals are filled by 
contributions from the K\"ahler potential.

These matrices exhibit alignment. 
To examine this, we describe
the ``misalignment" by dimensionless
quantities $\delta$, which are defined as 
the relative sizes of the off-diagonal to main diagonal
entries of the squark mass matrices at the basis where Yukawa
matrices are diagonal \cite{FCNC}. 
These quantities are severely limited by 
FCNC ($K-{\overline K}$ mixing, in particular) 
and CP-violating processes. 

In our model, 
all of the off-diagonal elements 
of $\delta_{LR}$ matrices are negligibly small.
This is because 
the soft trilinear terms are generated from 
the second and third terms in the bracket of eq. (\ref{pot}),
and are proportional to the Yukawa matrices. 
The biunitary matrices which diagonalize the Yukawa matrices
also diagonalize the left-right scalar 
mass matrices ${\tilde m}^2_{LR}$. 
The diagonal elements $|(\delta_{11}^{l,d,u})_{LR}|$ 
are also small, compared to that of 
general supergravity models, because they are suppressed
by extra powers of $\lambda$ and heavier squark masses.
The constraints from electric dipole moments 
of $u$, $d$ quarks and electron 
are thus easily satisfied \cite{Dva1, Moha}.

Now we examine the limits for 
$\delta_{LL}, \delta_{RR}$.
The most severe constraint 
comes from $K-{\overline K}$ mixing.
Following Ref. \cite{FCNC}, we deduce the limits 
\begin{equation}
|\sqrt {{\rm Re}(\delta^d_{12})^2_{LL}}|,~ |\sqrt {{\rm Re}(\delta^d_{12})^2_{RR}}| 
< 3.5 \times 10^{-2}, \ \ 
|\sqrt {{\rm Re}(\delta^d_{12})_{LL}(\delta^d_{12})_{RR}}| 
< 1.7 \times 10^{-2} ,
\end{equation}
where we take ${\tilde m}_q\sim 2$ TeV, $m_{\tilde g}^2/{\tilde m}_q^2
\simeq 1.6\times 10^{-3}$.
In our model, we find $|(\delta_{12}^d)_{LL}|\sim\lambda^7$
and $|(\delta_{12}^d)_{RR}|\sim\lambda^9$, and are much smaller 
than the above bounds. Other bounds from $\mu\to e\gamma$ and 
$D-{\overline D}$ and $B-{\overline B}$ mixings are also satisfied. 
The smallness
of these $\delta$ parameters ensure that 
the limits from CP-violating processes,
{\it e.g.}, CP-violating phases $\epsilon, \epsilon'$, are
met as well. Therefore this model poses no FCNC and CP problems.

\noindent {\bf 4}. For several anomalous $U(1)$ symmetries,
there exists a unique linear combination which is anomalous,
while the other combinations of 
$U(1)$'s are anomaly free \cite{string}.
In our model, 
the anomalous $U(1)$ and anomaly-free 
$U'(1)$ charges are related to the $U(1)_1\times U(1)_2$
charges by,
\begin{eqnarray}
&& q = (q_1 \delta^1_{\rm GS} + q_2 \delta^2_{\rm GS})/
{\sqrt {(\delta^1_{\rm GS})^2 + (\delta^2_{\rm GS})^2}}
= {4\over 5} q_1 + {3\over 5} q_2, \nonumber\\
&& q' = (q_1 \delta^2_{\rm GS} - q_2 \delta^1_{\rm GS})/
{\sqrt {(\delta^1_{\rm GS})^2 + (\delta^2_{\rm GS})^2}}
= {3 \over 5} q_1 - {4\over 5} q_2\ .
\end{eqnarray}
We see that $U'(1)$ charges 
of the superfields 
are flavor dependent. How these $U(1)$ symmetries originate from 
an underlying theory remains an open question.

An unwelcome feature of our model may be the $\mu$-problem.
Because of the symmetry, the leading contribution to the
$\mu$-term is of order $\lambda^4 m \simeq 5 $ GeV, which is too small.
On the other hand, the leading contribution to the $B\mu$-term 
is of order $\lambda^2 m\mpl$, and destablizes the hierarchy.

One can introduce a discrete symmetry to solve these problems.
We consider for example a $Z_3$ symmetry, under 
which all the standard model fields have charge one and the singlets
$\phi_{\pm}, \psi_{\pm}$ have charge zero. In addition, we introduce
a gauge singlet $X$, which has 
$Z_3$ charge one. Then the dangerous $B\mu$-term is not allowed;
instead one finds the following $Z_3$ invariant 
terms in the superpotential,
\begin{equation}
W \supset X^3 + X H_1 H_2 .
\label{eq:mu}
\end{equation}
If the $X$ field gets a negative soft mass $\sim -\lambda^2 m^2$ from  
gravitational interactions, a $\mu$-term of order $\lambda m$ is 
generated from eq. (\ref{eq:mu}). 
The soft $B$-term is of order $\lambda^2 m$.

Note that the model has a hierarchial mass structure, 
${\tilde m}_{Q_3,U_3,L_3}, m_{H_1,H_2} \simeq \lambda m$,
while all other sfermions have TeV masses. (
The heavy right-handed bottom squark does not 
give rise to a fine tuning problem because $\tan\beta$ and $\lambda_b$
are small.) The light fields have a 
low energy mass spectrum similar to 
that of supergravity models.
For example, $m = 2$ TeV in this model gives soft parameters
${\tilde m}_{Q_3, U_3, L_3}, m_{H_1, H_2}\simeq$ 400 GeV, 
$m_\lambda\simeq$ 80 GeV,
an $A$-term of about 80 GeV and $\tan\beta \simeq 2$. It gives the same
${\tilde t}_{1,2}$, ${\tilde b}_L,\ {\tilde \tau}_L$, ${\tilde\nu_\tau}$,
gaugino, Higgisino and Higgs masses as a supergravity model
with GUT scale soft parameter $M_0 \simeq 400$ GeV and 
$M_{1/2}\simeq $ 80 GeV. (This is because the first two generation
soft parameters decouple from the renormalization group equations for 
the third generation soft parameters and Higgs scalar masses.) 
The lightest colored particle is the gluino, and only 
light gauginos are reachable for the next generation 
linear $e^+ e^-$ collider, where the signals for light 
chargino and neutralino 
production are similar for the anomalous $U(1)$ and supergravity 
scenarios.

\noindent {\bf 5}.
To summarize, we here introduced a model based on an 
anomalous $U(1)_1\times U(1)_2$
symmetry, where supersymmetry breaking 
is induced by a Fayet-Iliopoulos $D$-term. 
This Abelian symmetry plays the role of a horizontal flavor
symmetry which gives a correct Yukawa texture. 
The quark and squark mass matrices are aligned, 
and thus the contributions to 
FCNC and CP-violating process are highly suppressed.

I would like to thank Jonathan Bagger and Chris Kolda for helpful discussions. 
This work was supported by the U.S. 
National Science Foundation, grant NSF-PHY-9404057. 




\bigskip


\bibliographystyle{unsrt}

\end{document}